\documentstyle[epsf]{mn} 

\onecolumn
 
\def\kms{km ${\rm s}^{-1}$\ }
\def\gx339{GX~339$-$4}

\title{Optical Spectroscopy of GX~339$-$4 during the High-Soft and 
   Low-Hard States} 

\author[R. Soria et al.]{Roberto Soria$^1$, Kinwah Wu$^2$, 
                         Helen M. Johnston$^3$ \\
$^1$ Research School of Astronomy and Astrophysics, Australian 
   National University, Private Bag, Weston Creek Post Office, \\ 
\ \  ACT 2611, Australia; roberto@mso.anu.edu.au \\
$^2$ Research Centre for Theoretical Astrophysics, School of Physics,
   University of Sydney, NSW 2006, Australia; \\ 
\ \  kinwah@physics.usyd.edu.au \\
$^3$ Anglo-Australian Observatory, P. O. Box 196, Epping, NSW 1710, 
   Australia; hmj@aaoepp.aao.gov.au  }

\date{Received: } 

\begin{document}

\maketitle

\begin{abstract} 
We carried out spectroscopic observations of the candidate black hole 
binary \gx339 during its low-hard and high-soft X-ray states. We have 
found that the spectrum is dominated by emission lines of neutral 
elements with asymmetric, round-topped profiles in the low-hard state. 
In the high-soft state, however, the emission lines from both neutral 
and ionised elements have unambiguously resolved double-peaked profiles. 
The detection of double-peaked emission lines in the high-soft state 
with a larger peak separation for higher-ionisation lines indicates 
the presence of an irradiatively-heated accretion disk. The round-topped 
lines in the low-hard state is probably due to a dense matter outflow 
from an inflated non-Keplerian accretion disk. Our data do not show velocity 
modulations of the line centres due to the orbital motion of the 
compact object, neither do the line base-widths show substantial  
variations in each observational epoch. There are no detectable 
absorption lines from the companion star. All these features are consistent 
with those of a system with a low-mass companion star and low orbital 
inclination.  

\end{abstract}


\section{Introduction}   

Almost all known black hole candidates (BHCs) in our Galaxy are
X-ray transients. Although the triggering for the X-ray outbursts is
not fully understood, models invoking accretion disk instability seem
to provide an explanation. Other important issues concerning the BHCs
are the transition between the X-ray spectral states and its relation
to the physical conditions at the accretion disk.

\gx339 is an X-ray transient (Markert et\,al.\ 1973; 
Harmon et\,al.\ 1994), and is also a radio source 
(e.g.\ Hannikainen et\,al.\ 1998) which shows jet-like features 
(Fender et\,al.\ 1997). It is classified as a BHC, because of its 
short-term X-ray and optical variability (Makishima et\,al.\ 1986), 
the transition between high-soft and low-hard X-ray spectral states 
(Markert et\,al.\ 1973), and the extended high energy power-law tail 
in its X-ray spectrum (Rubin et\,al. 1998). In fact it is also one of 
the few BHCs that have shown four X-ray spectral states: {\it off, 
low-hard, high-soft and ultra-high}. 
Its off state is generally characterised by very weak, hard X-ray 
emission. The optical counterpart is faint with 
$V \sim 19$--21. In the low-hard state, the 
source has a very hard X-ray spectrum, with an extended power-law 
component with a photon index $\sim 1.5$. The 2--10~keV X-ray 
flux is about 
$\sim 0.4 \times 10^{-9}\;{\rm erg\,cm^{-2}\,s^{-1}}$. The optical 
brightness is $V \sim 16$.  In the high-soft state, the 
X-ray spectrum is dominated by a soft thermal component. The 
power-law tail is weaker than that in the low-hard state and has 
a photon index $\sim 2$. The 2$-$10~keV X-ray flux is about 20 times  
higher than that of the low-hard state. The 
optical brightness is $V \sim 16$, similar to that in the low-hard 
state. In the ultra-high state, both the thermal and the power-law 
components are very strong in the X-ray spectrum. The 2$-$10~keV 
X-ray flux is about 50 times the X-ray flux in the low state, and 
the photon index of the power law is $\sim 2.5$. On one occasion 
Mendez \& van der Klis (1997) reported that the source was in a 
state intermediate between the 
low-hard and the high-soft state in which the 2$-$10~keV X-ray flux 
is about a factor of 5 below the X-ray flux in the high state.   
(For reviews of the optical and X-ray properties of the system, see 
e.g.\ Motch et\,al.\ 1985; Ilovaisky et\,al.\ 1986; 
Makishima et\,al.\ 1986; Corbet et\,al.\ 1987; Tanaka \& Lewin 1995; 
Mendez \& van der Klis 1997). 
 
Despite the fact that \gx339 is optically bright ($V\sim 16$) 
when it is X-ray active, it is not well studied in the optical 
bands. Its mass function has not yet  been determined, and so the  
black hole candidacy is not verified in terms of the orbital 
dynamics. From a photometric observation during the off state, 
Callanan et\,al.\ (1992) detected a periodicity of 14.8~hr, and  
attributed it to the orbital period. If this is true, the 
companion star would be a late-type star with a mass 
$< 1.6$~M$_\odot$.  

Optical spectra taken by Smith, Filippenko \& Leonard (1999) 
in 1996 May, when the system was in a low-hard state,
show a strong H$\alpha$ emission 
line with a broad flat-topped profile. Within the resolution and 
the signal-to-noise ratio limit of the spectra, the line resembles 
the double-peaked lines that characterise the accretion disk in binary 
systems. By fitting two gaussians to the H$\alpha$ line 
profile, a velocity separation of $370\pm40$ \kms is deduced. 
As the separation of the peak in this data set is not as clearly 
seen as in the other BHCs, e.g., GRO~J1655$-$40 
(Soria, Wu \& Hunstead 1999) and A0620$-$00 
(Johnston, Kulkarni \& Oke 1989), there is a  
possibility that the flat-topped profile seen in 1996 May is not 
intrinsically double-peaked, just as in the case of 
Cyg X-1 (Smith, Filippenko \& Leonard 1999). 

Here we report spectroscopic observations of \gx339 which we carried
out in 1997 May and in 1998 April and August. The system was in a
low-hard state in 1997 May and was in a high-soft state in 1998 April
and August (see Fig.~1). During the 1998 August observations, we also
obtained simultaneous photometric data. Our data show that \gx339 has
distinct optical spectral features in the low-hard and in 
the high-soft states, an indication of different physical conditions
in the line emission regions or perhaps different line emission
regions in the two X-ray spectral states.

\begin{figure}
\vspace*{0.75cm}
\begin{center}
\epsfxsize=12.5cm 
\epsfbox{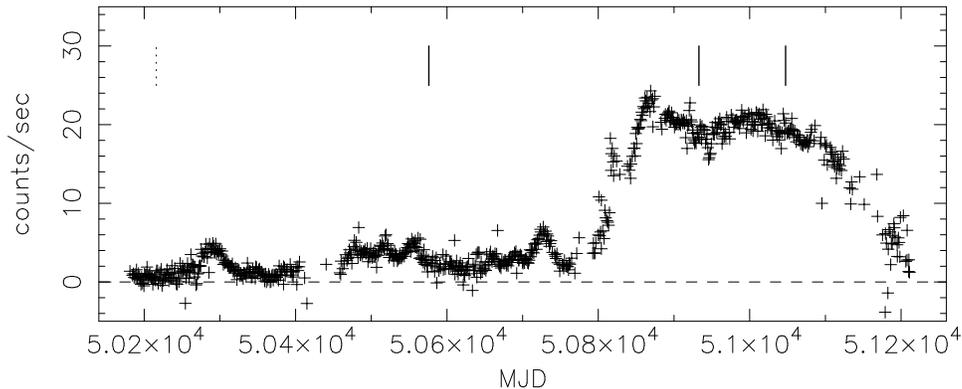} 
\end{center}
\caption{The RXTE/ASM light curve of \gx339 during the low-hard 
 state and the high-soft state between 1996 and 1998. The dates at 
 which our observations were carried out are marked by solid 
 vertical lines and that of the observation by 
 Smith et\,al.\ (1999) is marked by a dotted vertical line. }
\end{figure} 

\section{Observations and Data Reduction}

\begin{table*}
 \centering
 \begin{minipage}{140mm}
  \caption{Log of our spectroscopic observations of \gx339.}
  \begin{tabular}{lccc}
\hline
\hline
Date & HJD range  & Wavelength range & Resolution \\
& (HJD - 2450000)  &  (\AA) 
& (\AA\ FWHM) \\
\hline
\hline
\multicolumn{4}{c}{AAT 3.9m} \\
\hline
1997 May 6 & 574.966--575.226 & 5355--6950 & 3 \\
1997 May 8 & 576.974--577.042 & 5355--6950 & 3 \\
\hline
\multicolumn{4}{c}{ANU 2.3m} \\
\hline
1998 April 28 & 931.953--932.318 & 4150--5115 & 1.3 \\
&&6200--7150 & 1.3 \\
1998 April 29 & 932.943--933.268 & 4150--5115 & 1.3 \\
&&6200--7150 & 1.3 \\
1998 April 30 & 934.106--934.233 & 4150--5115 & 1.3 \\
&&6200--7150 & 1.3 \\
1998 August 20 & 1045.864--1045.140 & 4150--5115 & 1.3 \\
&&6200--7150 & 1.3 \\
1998 August 23 & 1048.860--1049.103 & 4150--5115 & 2 \\
&&6200--7150 & 2 \\  
\hline
\end{tabular}
\end{minipage}
\end{table*}

The 1997 May 6 and 8 observations of \gx339 were carried out  with the 
RGO spectrograph and Tektronix 1k$\times$1k thinned CCD on the 
3.9\,m Anglo-Australian Telescope (AAT). 
The seeing condition was 2~arcsec. Series of 600~sec 
spectra were taken in the $5355-6950$ \AA\ region. 
A grating with 300 grooves/mm was used with the 25~cm camera, giving
a resolution of 3 \AA\ FWHM.   

On 1998 April 28 $-$ 30 we observed the system again, with 
the Double Beam Spectrograph (DBS) on the ANU 2.3\,m Telescope 
at Siding Spring Observatory. 
The detectors on the two arms of the spectrograph 
were SITe 1752$\times$532 CCDs. Gratings with 1200 grooves/mm were used 
for both the blue ($4150-5115$ \AA) and the red ($6200-7150$ \AA) bands,
giving a resolution of $1.3$ \AA\ FWHM. The average seeing on each 
of the three nights was about 2 arcsec. Further observations 
were carried out on 1998 
August 20 and 23, with the same instrumental setup as in the April 
observations.  Simultaneous photometric observation were conducted with 
the ANU 40\,in telescope on 1998 August 20 and 23. The seeing 
was 2 and 3 arcsec on the two nights respectively. 
A log of our spectroscopic observations is shown in Table 1.

Standard data reduction procedures were followed, using the IRAF tasks. 
After removing the bias and pixel-to-pixel gain variations from each 
exposure, we subtracted the sky background and extracted the spectra 
with the APALL routine. CuAr and FeAr lamp spectra 
were taken to carry out wavelength calibration. We removed 
atmospheric spectral features using the standard star LTT 7379.

\section{Results}  

\subsection{Spectrum in the low-hard state}    

 \begin{figure}
\vspace*{0.75cm}
\begin{center}
\epsfxsize=12.5cm 
\epsfbox{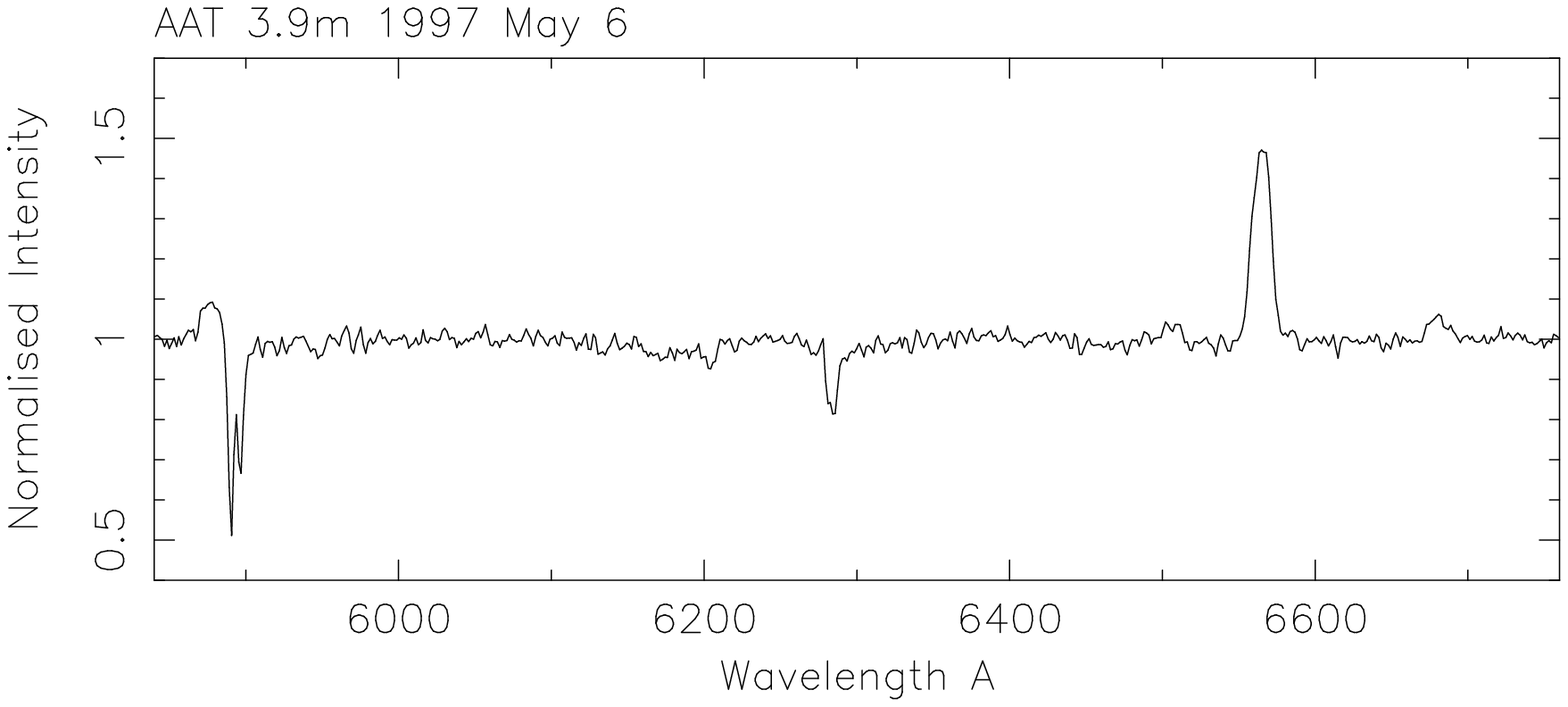} 
\end{center}
\caption{The summed spectra of \gx339 at the H$\alpha$ region. 
   The data were obtained by the AAT on 1997 May 6, when the system 
   was in a low-hard X-ray state. The spectrum is normalised to the 
   continuum, and wavelengths are vacuum, heliocentric. }
\vspace*{0.6cm}
\begin{center}
\epsfxsize=12.5cm 
\epsfbox{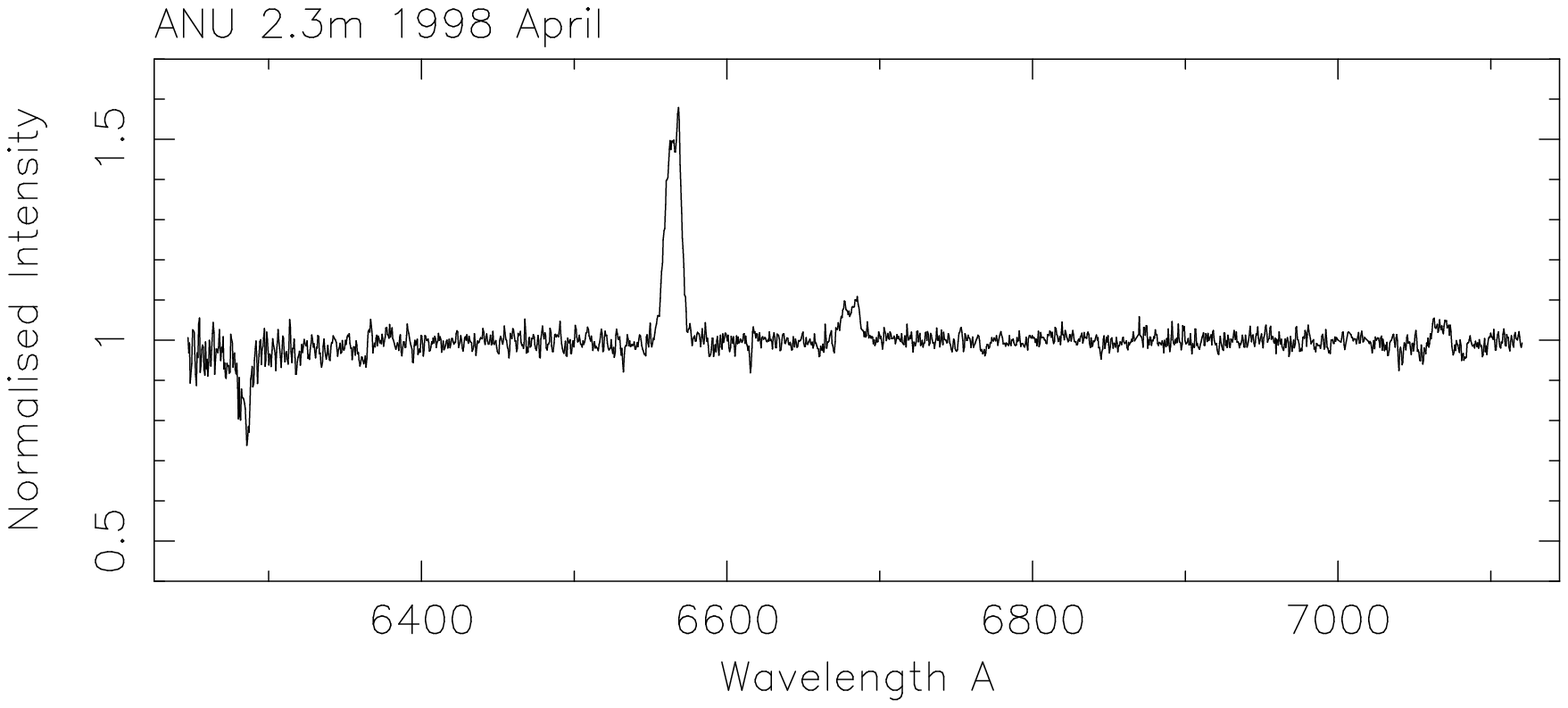} 
\end{center}
\caption{The summed spectra of \gx339 in the H$\alpha$ region with 
   the same scale as in Fig.~2. The data were obtained by the 
   ANU 2.3\,m telescope on 1998 April 28 $-$ 30, when the system 
   was in a high-soft X-ray state. }
\vspace*{0.6cm}
\begin{center}
\epsfxsize=12.5cm 
\epsfbox{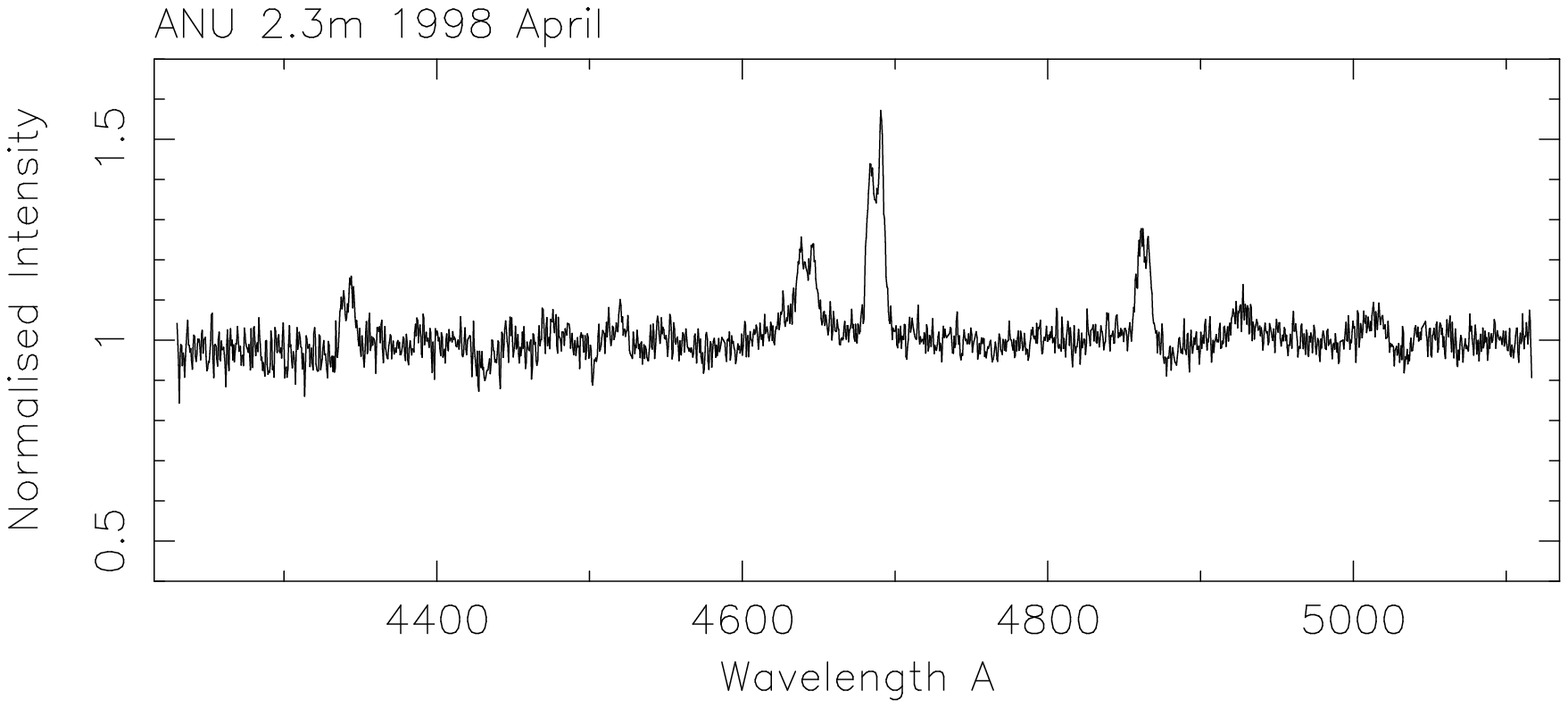} 
\end{center}
\caption{Same as Fig.~3 for the H$\beta$/He\,{\scriptsize II} 
   region. }
\end{figure} 

Figure 2 shows the summed spectrum from the observations we conducted 
on 1997 May 6. The total exposure time was 19,800 sec. The H$\alpha$, 
He\,{\scriptsize I} $\lambda$\,6678 and 
He\,{\scriptsize I} $\lambda$\,5876 emission lines are easily identified. 
The emission feature at about 6500 \AA\ is probably the 
N\,{\scriptsize II} $\lambda$\,6505 emission line.  
(The main features identified in the spectra and their equivalent 
widths are listed in Table 2.) Our data do not reveal 
two clearly resolved peaks for H$\alpha$, 
He\,{\scriptsize I} $\lambda$\,5876 and 
He\,{\scriptsize I} $\lambda$\,6678. 
Although the suspected N\,{\scriptsize II} $\lambda$\,6505 line appears 
to have two maxima, it is uncertain whether they are purely statistical 
or intrinsic. We do not detect the 
Li\,{\scriptsize I} $\lambda$\,6708 absorption line, which was present 
in the spectra of the BHCs 
GRO~J1655$-$40, A~0620$-$00, V404~Cyg and Cen X-4 
(Martin et\,al.\ 1992; Shahbaz et\,al.\ 1999; Smith et\,al.\ 1999), nor 
any other obvious stellar absorption lines. All absorption features 
in our spectra appear to be interstellar.

The H$\alpha$ line is asymmetric, with its peak slightly skewed 
towards the red. The line profile is more appropriately described 
as asymmetric and round-topped rather than as unevenly double-peaked. 
The equivalent width (EW) of the line is $-7.2 \pm 0.3$ \AA\ (negative values 
are taken to indicate emission). The EW and the overall shape are 
similar to those found in the spectra obtained by 
Smith et\,al.\ (1999) on 1996 May 12, when the system was also in a 
low-hard state, except for the fact that in those observations the 
line peak was skewed towards the blue instead. We considered a 
two-gaussian fit using the QDP routine (Tennant 1991) and obtained 
the gaussian centres at 6560.15 and 6567.57 \AA. If 
we assume that the line truly has a double-peaked profile, the 
deconvolved peak separation is then roughly 7.4 \AA. This value is 
consistent with the $8.0\pm0.8$ \AA\ separation obtained by 
Smith et\,al.\ (1999) with a two-gaussian fit to their 1996 data.    

\subsection{Spectrum in the high-soft state}  

In Figure 3 we show the summed spectrum in the H$\alpha$ region for 
the data obtained on 1998 April 28 $-$ 30; in Figure 4, we show the 
summed spectrum in the the H$\beta$/He\,{\scriptsize II} region for 
the data obtained on the same nights. The total exposure times in 
both case were 25,500 sec on the first night, 19,500 sec on the second 
and 4,000 on the third. Strong H\,{\scriptsize I} Balmer emission lines 
are seen in the spectra. Other prominent emission lines are 
He\,{\scriptsize I} $\lambda$\,6678, 
He\,{\scriptsize I} $\lambda$\,7065, 
He\,{\scriptsize II} $\lambda$\,4686 
and N\,{\scriptsize III} $\lambda \lambda$\,4641,4642. 

The H$\alpha$, H$\beta$, H$\gamma$, 
He\,{\scriptsize I} $\lambda$\,6678, 
He\,{\scriptsize II} $\lambda$\,4686 and 
N\,{\scriptsize III} $\lambda \lambda$\,4641,4642 
lines all show double-peaked profiles, and the peaks are resolved 
visually. The peaks are still clearly separated even when we 
bin the data to mimic a spectrum with a resolution of 4 \AA\ FWHM, 
lower than the 3 \AA\ resolution of our 1997 May spectra 
(see Fig.~2, cf.\ also Smith et\,al.\ 1999). It is therefore 
the first time that double-peaked lines are unambiguously detected 
in the optical spectra of \gx339. The peak separations of the
emission lines are listed in Table 3.

We note that while H$\alpha$ 
and H$\beta$ has similar velocity separation, that of H$\gamma$ is 
significantly larger. As the signal-to-noise ratio of the H$\gamma$ line 
is relatively low, the unexpectedly large velocity separation is  
probably caused by contamination.  

The general features of the 1998 August spectra (not shown) are similar 
to those in the 1998 April spectra. The H$\alpha$, H$\beta$, H$\gamma$, He\,{\scriptsize I} $\lambda$\,6678, 
He\,{\scriptsize II} $\lambda$\,4686 and 
N\,{\scriptsize III} $\lambda \lambda$\,4641,4642 lines are prominently 
seen in emission and generally have double-peaked profiles. Because of 
poorer  observing conditions -- cloudy nights -- the peaks are not as 
well resolved as those appearing in the 1998 April spectra.  

\begin{table*}
\centering
\begin{minipage}{140mm}
\caption{ The equivalent widths (in \AA) of the most prominent 
   emission lines  
   from \gx339 and the interstellar absorption lines. }
  \begin{tabular}{lccc}
\hline
\hline
   & 1997 May & 1998 Apr & 1998 Aug \\
\hline
\hline
\multicolumn{4}{c}{emission lines from \gx339}  \\
\hline
H$\gamma$ $\lambda$\,4340 
         &  & $-2.0$ & $-1.8$ \\
N\,{\scriptsize III} $\lambda \lambda$\,4641,4642 
         &  & $-2.5$ & $-3.0$ \\
He\,{\scriptsize II} $\lambda$\,4686 
         &  & $-5.3$ & $-5.1$ \\ 
H$\beta$ $\lambda$\,4661 
         &  & $-3.0$ & $-3.5$ \\ 
He\,{\scriptsize I} $\lambda$\,4922		
         &  & $-0.9$ & $-1.2$ \\ 
He\,{\scriptsize I} $\lambda$\,5876 
         & $-1.8$ &  &     \\ 	
N\,{\scriptsize II} $\lambda$\,6505 
         & $-0.7$ &  ***  &  \\ 
H$\alpha$ $\lambda$\,6563 
         & $-7.2$ & $-6.7$ & $-7.4$ \\ 
He\,{\scriptsize I} $\lambda$\,6678 
         & $-1.0$	& $-1.3$ & $-1.3$ \\ 
He\,{\scriptsize I} $\lambda$\,7065	
         &  & $-1.0$ & $-1.0$ \\ 	
\hline
\multicolumn{4}{c}{interstellar (IS) lines {$^{\dag}$}} \\ 
\hline 
$\lambda$\,4428	 &     & 0.7  & 0.6 \\ 
$\lambda$\,4502  &     & 0.3  & 0.4 \\ 
$\lambda$\,4726	 &     &  ***  & 0.4 \\ 
$\lambda \lambda$\,5778,5780	
                 & 1.1 &      &     \\ 	
NaD{\scriptsize 2} $\lambda$\,5890	
                 & 2.1 &      &     \\ 	
NaD{\scriptsize 1} $\lambda$\,5896	
                 & 1.6 &      &     \\ 
$\lambda$\,6202	 & 0.6 & 0.3  & 0.3 \\ 
$\lambda$\,6270	 & 0.2 & 0.2  & 0.2 \\ 
$\lambda$\,6284	 & 1.8 & 1.7  & 1.6 \\ 
$\lambda$\,6613	 & 0.2 & 0.2  & 0.2 \\ 
\hline 
\end{tabular}  
\begin{tabular}{lc}
{\footnotesize {***} too weak to be accurately measured } &  \\ 
{\footnotesize {$^{\dag}$\ \ \ } classified as IS lines according   
   to Herbig (1975) }  & \\ 
\end{tabular}    
\end{minipage}
\end{table*}

\begin{table*}
\centering
\begin{minipage}{140mm}
\caption{ The peak-to-peak velocity separation of the double-peaked 
   emission lines from \gx339 in the summed spectrum from 1998 April, 
   with a $2\sigma$ uncertainty estimated from the individual spectra.}
  \begin{tabular}{lc}
\hline
\hline
    Line & $\Delta v$ (\kms) \\
\hline
\hline
H$\alpha$ $\lambda$\,6563 
         & $250 \pm 20$ \\
H$\beta$ $\lambda$\,4861 
         &  $260 \pm 30$ \\
H$\gamma$ $\lambda$\,4340 
         &  $350 \pm 30$ \\
He\,{\scriptsize I} $\lambda$\,6678 
         & $380 \pm 20$\\ 
He\,{\scriptsize I} $\lambda$\,7065	
         & $390 \pm 20$ \\
N\,{\scriptsize III} $\lambda \lambda$\,4641,4642 
         & $490 \pm 15$ \\
He\,{\scriptsize II} $\lambda$\,4686 
         & $510 \pm 15$ \\ 
\hline 
\end{tabular}    
\end{minipage}
\end{table*}

\subsection{Line profile morphology}   

\begin{figure}
\vspace*{0.75cm}
\begin{center}
\epsfxsize=10cm 
\epsfbox{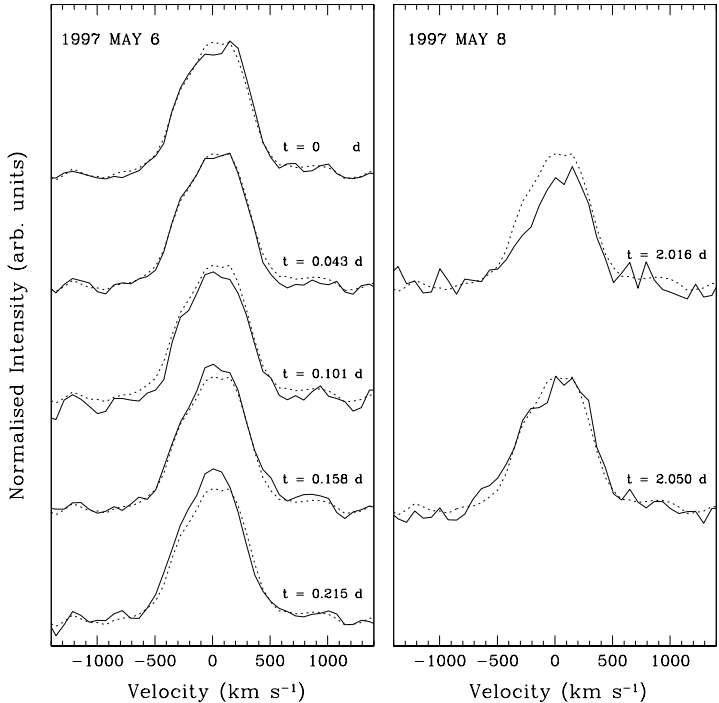} 
\end{center}
\caption{H$\alpha$ line profiles for the 1997 May 6 and 8 observations. 
  The zero velocity is defined by the wavelength of the line in its rest 
  frame. The continuum is normalised to unity. The time reference is 
  chosen such that ``$t=0$'' corresponds to the mid-time of the first 
  series of observations on May 6. The total integration times for each 
  of the combined profiles plotted here from top to bottom 
  are 2400, 3600, 4800, 4200 and 4800 s for the May 6 spectra, 
  and 3600 and 1800 s for the May 8 spectra.
  The averaged line profile for all the 1997 May 
  observations is shown as a dotted line for comparison. }
\end{figure} 

\begin{figure}
\vspace*{0.75cm}
\begin{center}
\epsfxsize=10cm 
\epsfbox{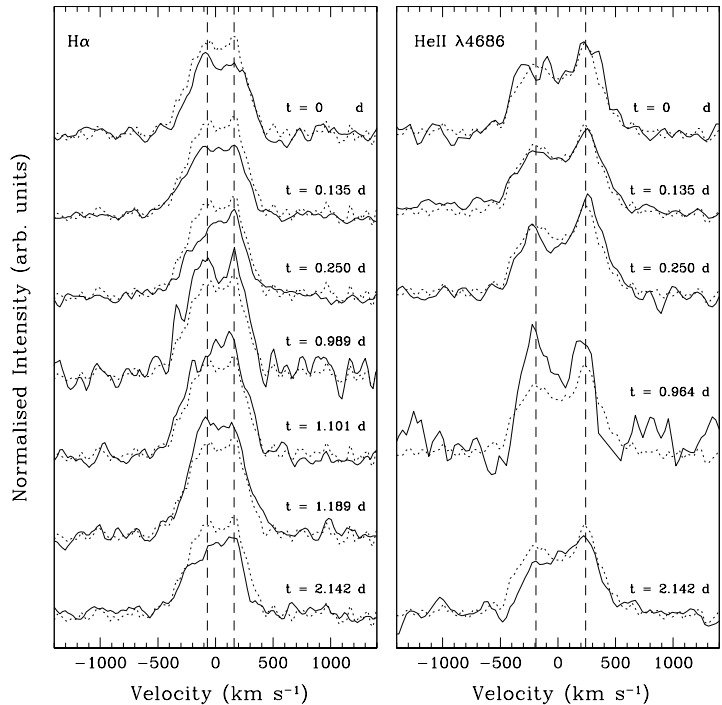} 
\end{center}
\caption{The profiles of the H$\alpha$ and the 
  He\,{\scriptsize II} $\lambda$\,4686 lines in the 1998 April 
  observations, with the zero velocity and continuum normalisation 
  defined as in Fig.~5.  The time reference ``$t=0$'' corresponds to 
  the mid-point of the first observation on April 28. The averaged 
  line profiles (dotted lines) and the locations of the peaks 
  (vertical dashed lines) are shown for comparison. }
\end{figure}  

\begin{figure}
\vspace*{0.75cm}
\begin{center}
\epsfxsize=10cm 
\epsfbox{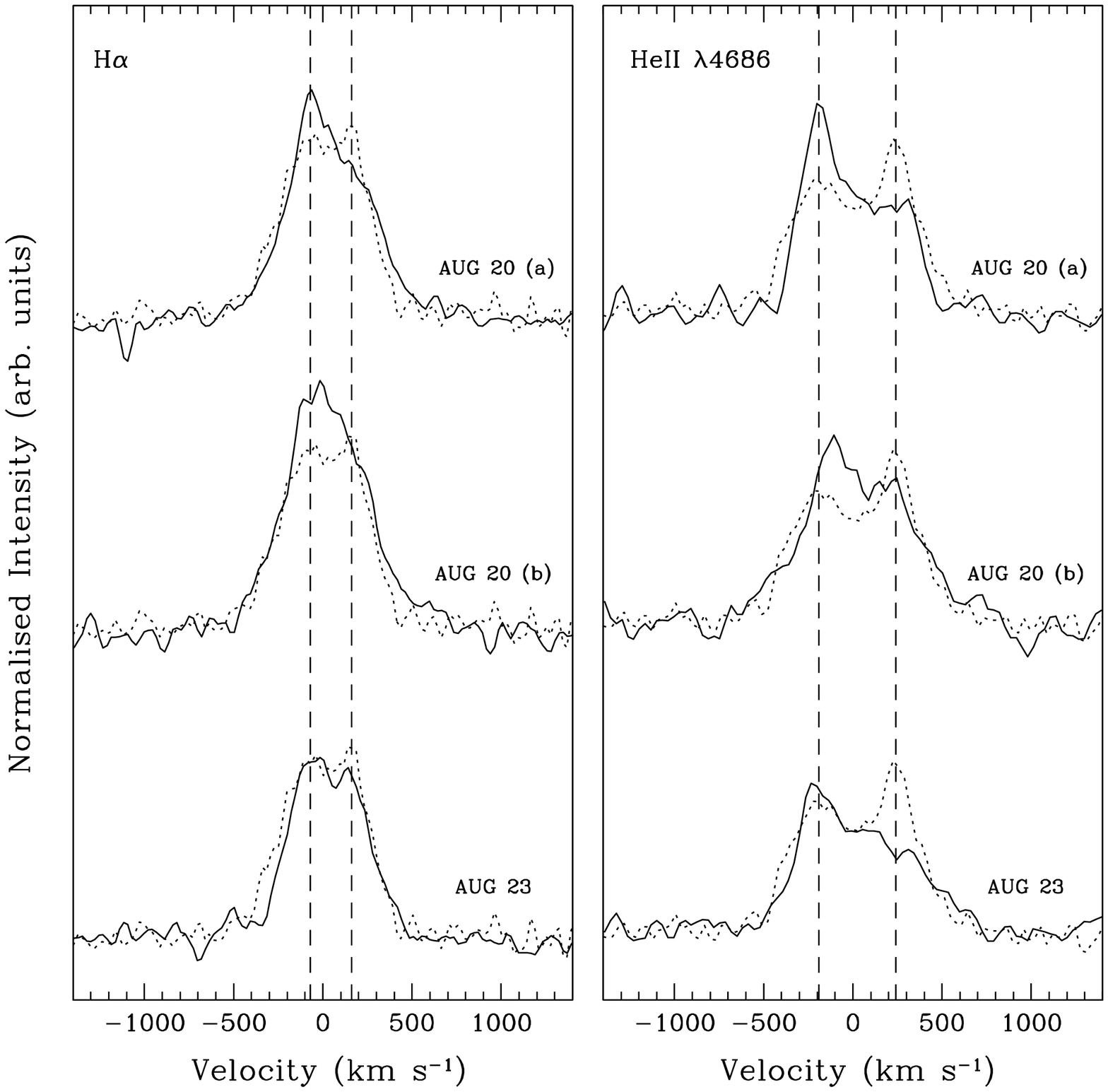} 
\end{center}
\caption{Same as Fig.~6 for the 1998 August 20 and 23 observations. 
  The averaged profile from the first half of the August 20 night 
  is labelled with ``AUG 20 (a)'', and the one from the second half, 
  ``AUG 20 (b)''. The time separation between the mid-points of the 
  two series of spectra is about 4 h. 
  The averaged line profiles of the 1998 April observations (dotted 
  lines) and the corresponding locations of the peaks (vertical 
  dashed lines) are shown for comparison. }
\end{figure}

In Figure 5, we show a sequence of H$\alpha$ line profiles from 
the 1997 May 6 and 8 observations. The H$\alpha$ line profiles are 
asymmetric, with a slightly red-shifted peak and a round top. There 
is no strong indication of double-peaked profiles in any set of data 
obtained in 
1997 May (cf.\ the H$\alpha$ and the He\,{\scriptsize II} 4686 lines 
in Fig.~6, where the double-peaked structure is instead very clear). 
The line shows detectable variations from night to night. If the 
line profiles are fitted with a single gaussian, one may obtain 
a series of velocity shifts. Judging from the central location of 
the line bases, we argue that such variations are statistical, 
i.e.\ our data do not show systematic velocity modulation due to 
orbital motion. The base-width of the line does not show obvious 
variations over the two nights of observations, either. The 
variations in the line top are therefore likely to be due to opacity 
rather than kinematic effects.  

The signal-to-noise ratios of the 
He\,{\scriptsize I} $\lambda$\,6678 line 
profiles in the 1997 May observations (not shown) are not as good as 
those measured at H$\alpha$. In spite of this, the properties of the 
He\,{\scriptsize I} $\lambda$\,6678 line profile sequence is still 
consistent with our ``argument'' of no 
systematic velocity modulations and no line-width variations.  

The sequence of H$\alpha$ and  
He\,{\scriptsize II} $\lambda$\,4686 line profiles from the 
1998 April observations are shown in Figure~6. The presence of a 
double-peak in each profile is not as obvious as it appears in the 
summed spectra. However, the trace of two peaks can still be easily 
seen. The relative prominence of the two peaks varies from one 
spectrum to another. As changes of the relative contribution of the 
two peaks have also been observed in other BHCs, e.g.\ GRO J1665$-$40,  
when they were in a high-soft state (see Soria, Wu \& Hunstead 1999), 
this may imply that similar activities are occurring in the accretion 
discs of these BHCs during that state.  

The profile of the  H$\alpha$ line is slightly skewed, with an 
apparently wider blue wing. We suspect that the skewness is caused 
by the presence of a weak broad component whose red wing is 
partially obscured/absorbed along the line of sight, an indication 
that opacity effects play a significant role in the high-soft state 
as well as in the low-hard state.  A hint of a third peak 
superimposed on the blue wing is detected in some spectra, 
particularly in the last series of observations from April 28. The 
profile sequence in Figure~5 reveals that the strength of the 
emission above the continuum increased from the first to the second 
night, and then decreased, to the previous levels, from the second 
to the third night. The average EW of the H$\alpha$ line was 
$-6.6 \pm 0.3$ \AA\ on April 28, $-9.0 \pm 0.2$ \AA\ on April 29, 
and $-6.0 \pm 0.3$ \AA\ on April 30. We do not have photometric data 
from 1998 April, and we are therefore unable to determine whether 
the variations in the EW are a consequence of changes in the intensity 
of the continuum or of the emission line itself.  

The presence of two peaks in the He\,{\scriptsize II} $\lambda$\,4686 
line is more obvious than in the case of H$\alpha$. They are clearly 
separated most of the time, and their separation is wider than that 
of the two peaks in the H$\alpha$ line profile. The red and blue wings 
are more symmetric with respect to each other, in comparison with the 
wings of the H$\alpha$ line.  The average EW of 
He\,{\scriptsize II} $\lambda$\,4686 was $-5.5 \pm 0.2$ \AA\ on April 28, 
$-6.0 \pm 0.4$ \AA\ on April 29, and $-4.5 \pm 0.3$ \AA\ on April 30.

In Figure 7 we show the profiles of the H$\alpha$ and the 
He\,{\scriptsize II} $\lambda$\,4686 lines from the 1998 August 
observations. The corresponding peak locations are consistent with 
those found in April. The red peak is generally weaker than that 
observed in 1998 April, while the blue peak seems to be stronger. 

\section{Discussion} 

\subsection{Accretion disk and line emission}   

The non-detection of Li\,{\scriptsize I} $\lambda$\,6708 and of other 
stellar absorption lines above the noise level (S/N $\sim 30$)
suggests that the emission from the companion star gives a negligible 
contribution to the optical continuum. This is not surprising: 
in 1981 the system was observed at an optical brightness as low as 
$V \sim 21$ mag (Hutchings, Cowley \& Crampton 1981; Ilovaisky 
\& Chevalier 1981), which is approximately the brightness of a 
$1 M_{\odot}$ main sequence star at a distance of $\sim 4$ kpc 
(Zdziarski et\,al.\ 1998); if that is the intrinsic brightness 
of the companion star, it would have contributed only $\sim 1/50$ of 
the optical flux detected in our observations ($V \sim 16.5$).

The spectra are dominated by emission lines in both the low-hard and 
the high-soft states. We argue that these lines are emitted from the 
accretion disc and/or its corona and wind. Thus, the accretion disc is 
present and active in both X-ray spectral states. 

The emission lines in the 1997 May spectra appear to have a
round-topped profile, skewed towards the red. Although we cannot rule 
out the interpretation that the round-topped lines actually consist  
of two components, we do not see clearly resolved peaks in our data. 
An alternative explanation is that the round-topped lines are 
the consequences of an outflow, a dense wind from an inflated 
non-Keplerian disk or an evaporating corona. If the outflowing matter
is sufficiently dense to produce a large opacity, lines emitted from 
the accretion disc beneath can be masked. The lines will then be 
dominated by emission from the outflowing material instead of 
by emission from the accretion disc. As the kinematic velocity
of the outflowing matter should be of the same order of magnitude as 
the local Keplerian velocities, the widths of the lines from the 
outflowing material may not differ much from the widths of lines 
from the accretion disc. However, the lines will not show  
double-peaked profiles, and will instead appear to be round-topped 
or flat topped, depending on the outflow velocity and density 
profiles (see Chapter 14 in Mihalas 1978), similar to those of 
lines from windy massive stars.    

The 1998 spectra show a strong optical continuum originated from an 
optically thick accretion disc. For a Keplerian disc, the rotational 
velocity and the temperature increase radially inward. If the disc is 
only viscously heated at the middle layer of the disc, a temperature 
gradient will be set up, such that the temperature decreases from 
the central plane to the disc surface. This will result in a 
spectrum dominated by absorption lines (see e.g.\ la Dous 1989). 
However, if there is a temperature inversion near the disc surface 
---  which may be caused by external irradiation ---  
the spectrum will then be dominated by emission lines. 
The 1998 April spectra clearly show prominent emission lines 
with resolved double-peaks superimposed on a strong continuum. 
This implies that the accretion disc is irradiatively heated. 
The increase in the velocity separation with the ionisation state 
of the accreting matter is strong evidence that the disc is  
irradiated by a central source, as the higher excitation lines 
originate from the inner 
regions, where the Keplerian velocities are larger and the 
temperature higher due to the smaller distance from the  
irradiation source. For a discussion on irradiation heating of accretion 
disks, see e.g. Dubus et\,al.\ (1999). 
 
The velocity separation of the H$\alpha$ peaks in the high-soft state 
is smaller than the velocity separation obtained by a two-gaussian 
fit to the line profiles obtained in the low-hard state 
(see also Smith et\,al.\ 1999). It is, however, roughly the same as 
the velocity separation of the H$\beta$ line observed in the same 
epoch. As the velocity separation of the peaks in the low-hard state 
is not well-defined, we cannot draw any firm conclusions from the 
comparison. However, we have found that the velocity separation of 
the peaks in the H$\alpha$ line in the high-soft state is a factor 
of two smaller than that observed in 
GRO J1655$-$40 in the same X-ray spectral state 
(Soria, Wu \& Hunstead 1999). The latter is a high orbital inclination 
system ($i \approx 70^o$, e.g.\ van der Hooft et\,al.\ 1998).  
If the masses of the black holes in \gx339 and GRO J1655$-$40 are 
similar, say 7--10\,M$_\odot$, the smaller velocity separation for 
\gx339 implies that either the accretion disc in \gx339 is larger or 
the orbital inclination of \gx339 is lower. As the companion star of 
\gx339 is not visible during its off state (Callanan et\,al.\ 1992), 
it is likely to be a low mass star, less massive than the companion star 
in GRO J1655$-$40. This leads us to propose that \gx339 is a system with 
a low orbital inclination. A low orbital inclination for \gx339 is in 
fact consistent with the null detection of the orbital modulation in 
our 1997/1998 spectroscopic data. It is also supported by the fact 
that the radio spectrum of \gx339 is relatively flat 
(Fender et\,al.\ 1997), unlike the steep spectra expected from a 
superluminal synchrotron jet source. 

Although a 14.8-h periodicity was detected when \gx339 was observed 
in an off state (Callanan et\,al.\ 1992), both our 1997/1998 
spectroscopic data and our 1998/1999 photometric data 
(Soria, Wu \& Johnston 1999a,b) do not show any obvious periodicities. 
Our photometric observations in 1998 August show variations in the 
brightness of the source similar to the flaring activities seen in 
a previous observation by Corbet et\,al.\ (1987). 
On August 20 its brightness varied between $V = 16.45 \pm 0.01$ and 
$V = 16.52 \pm 0.01$, while on August 23 the source brightened up from 
$V = 16.61 \pm 0.01$ at the beginning of the night to a maximum of 
$V = 16.22 \pm 0.01$ 5 hours later, then started to decline again.
Further photometric observations during the 
off state are required to verify if \gx339 has a 14.8-h 
orbital period. If we accept the argument that \gx339 has 
a low orbital inclination, the 14.8-h periodicity detected by 
Callanan et\,al.\ (1992) may perhaps also be due to the precession of a weak 
accretion disc and/or its associated jet. 

\subsection{Summary}  

We carried out spectroscopic observations of the BHC binary \gx339 
during its low-hard and high-soft states. Our data show that the 
optical spectra in the low-hard state are characterised by emission 
lines with slightly asymmetric, round-topped profiles; on the other hand, 
the spectra in the high-soft state show emission lines with unambiguously 
resolved doubled-peaked profiles. We do not see obvious stellar absorption 
lines: this implies that the optical spectrum in both X-ray spectral states 
is dominated by emission from accreting matter around the black hole. The 
round-topped lines seen in the low-hard state are probably formed in 
opaque matter outflowing from the central black hole and/or the 
accretion disc. The double-peaked lines seen in the high-soft state, 
however, indicate the presence of a bright, active accretion disc. The 
trend of increasing velocity separation with line ionisation 
supports the model of an accretion disc irradiatively heated by soft X-rays 
from a central source. We do not see 14.8-h modulations in our spectroscopic 
data and we are therefore unable to verify the 14.8-h orbital period. 
The null detection of the 
14.8-h periodicity and of stellar absorption lines, together with 
the relatively small velocity separation of the line peaks, lead us to 
believe that \gx339 is low-mass system with a low orbital inclination. 

\section{Acknowledgements} 

We thank Richard Hunstead for discussions and Michelle Buxton for taking 
the 1998 August 23 spectra. KW acknowledges the support 
from the Australian Research Council through an Australian Research 
Fellowship and an ARC grant.

\end{document}